\documentclass[journal]{IEEEtran}
\usepackage{booktabs}
\usepackage{graphicx}
\usepackage{listings}
\usepackage{xcolor} 
\usepackage[colorlinks=true, 
            linkcolor=blue,     
            citecolor=blue,     
            urlcolor=blue       
           ]{hyperref}

\begin{document}

\title{Beyond IVR Touch-Tones: \\Customer Intent Routing using LLMs}


\author{\IEEEauthorblockN{Sergio Rojas-Galeano} \\
\IEEEauthorblockA{\textit{Facultad de Ingeniería} \\
\textit{Universidad Distrital Francisco José de Caldas}\\
Bogotá, Colombia \\
srojas@udistrital.edu.co}
}

\maketitle

\begin{abstract}
Widespread frustration with rigid touch-tone Interactive Voice Response (IVR) systems for customer service underscores the need for more direct and intuitive language interaction. While speech technologies are necessary, the key challenge lies in routing intents from user phrasings to IVR menu paths, a task where Large Language Models (LLMs) show strong potential. Progress, however, is limited by data scarcity, as real IVR structures and interactions are often proprietary. We present a novel LLM-based methodology to address this gap. Using three distinct models, we synthesized a realistic 23-node IVR structure, generated 920 user intents (230 base and 690 augmented), and performed the routing task. We evaluate two prompt designs: descriptive hierarchical menus and flattened path representations, across both base and augmented datasets. Results show that flattened paths consistently yield higher accuracy, reaching 89.13\% on the base dataset compared to 81.30\% with the descriptive format, while augmentation introduces linguistic noise that slightly reduces performance. Confusion matrix analysis further suggests that low-performing routes may reflect not only model limitations but also redundancies in menu design. Overall, our findings demonstrate proof-of-concept that LLMs can enable IVR routing through a smoother, more seamless user experience---moving customer service one step ahead of touch-tone menus.
\end{abstract}


\section{Introduction}
\IEEEPARstart{T}{he} modern customer service landscape is increasingly shaped by automation, with Interactive Voice Response (IVR) systems acting as the entry point for millions of daily interactions. Originally intended to streamline call routing and reduce costs, traditional touch-tone IVRs have become a source of widespread user frustration. Callers often struggle with navigating complex menus, repeatedly pressing buttons, enduring long wait times, and failing to reach the right department. These pain points translate into inefficient call handling, longer agent interactions, and ultimately, reduced customer satisfaction. This underscores the growing need for a more natural, efficient, and user-friendly alternative.

This demand has driven interest in natural language-based call routing. While research in Natural Language Processing (NLP) has long explored conversational interfaces using rule-based and machine learning approaches, recent advances in Large Language Models (LLMs) offer a transformative opportunity. LLMs can interpret and generate human-like text with high accuracy, positioning them as central to the Natural Language Understanding (NLU) component of next-generation IVR systems. When integrated with speech recognition and synthesis, LLMs promise to move beyond the limitations of rigid menu trees and toward flexible, conversational interfaces.

However, evaluating LLMs for IVR call routing presents significant challenges, especially around data availability. Unlike many NLP tasks supported by large public datasets, IVR systems are typically proprietary. Their hierarchical menu structures are confidential, and real user interaction data is rarely accessible due to privacy and legal constraints. This lack of transparent, domain-specific data limits both development and rigorous evaluation of natural language call routing solutions.

To address these challenges, we propose a synthetic, LLM-driven experimental methodology that simulates a realistic IVR environment and enables controlled evaluation. Our pipeline uses three LLMs to construct the experimental framework. The first generates a 23-node IVR structure for a hypothetical telecom company. The second produces a dataset of 920 user intents, combining a base set of 230 distinct intents with 690 linguistically varied augmentations. The third model performs the routing task by matching each intent to the correct IVR path.

We compare routing performance under two context representations: a \textit{Descriptive IVR Menu} (a plain-text hierarchical description) and \textit{Flattened IVR Paths} (a list of terminal nodes identified by their Dual-Tone Multi-Frequency—DTMF—sequences). Our findings highlight how the structure and clarity of context input can significantly affect routing accuracy and offer practical insights for future implementations of LLM-based IVR systems.

\section{Related Work}

The transition from rigid touch-tone IVR systems to more natural, conversational interfaces has long been a goal in automated customer service. Early studies, such as \cite{suhm2002comparative}, demonstrated the advantages of natural language call routing over traditional DTMF systems in terms of user experience and efficiency. This effort has driven decades of research in NLP for intent recognition and dialogue management, spanning rule-based approaches, statistical models, and traditional machine learning techniques.

Recent advances in Large Language Models (LLMs) such as GPT-3, InstructGPT, and GPT-4 \cite{brown2020language, ouyang2022training, openai2023gpt4} have significantly advanced conversational AI. These models can interpret free-form user input with unprecedented fluency and contextual awareness. However, their application to IVR systems—particularly for directly mapping unstructured user complaints to structured IVR menu paths—remains largely unexplored. Existing work on IVR optimization often employs machine learning but stops short of using LLMs for the routing task itself. For example, \cite{ilk2020improving} and \cite{ts2018comparative} apply statistical models and classifiers to improve triage and routing accuracy based on IVR data. Similarly, \cite{singh2025ai, singh2025streamlining} discuss the promise of transformer-based models like BERT and GPT for enhancing intent detection and NLU in chatbots and IVR applications. Yet, none report concrete implementations where LLMs perform direct complaint-to-path mapping in a phone menu context.

Other contributions, such as \cite{kosherbay2024ai}, propose AI-enhanced IVR architectures incorporating speech recognition and NLP. These works often focus on system-level design or high-level integration strategies but do not provide empirical evaluations of LLMs applied to the routing task. A recent scientometric review by \cite{coman2025ivr} confirms this gap, surveying decades of IVR research without identifying any study that evaluates LLMs for path selection based on user complaints.

In summary, while prior research acknowledges the potential of advanced NLP to improve IVR routing, it lacks direct exploration of LLMs for mapping natural language input to structured IVR menu paths. This absence highlights the novelty and relevance of our proposed methodology.

\section{Methodology}

\subsection{Experimental Design Overview}

To address the data scarcity problem in developing and evaluating natural language IVR systems, we propose a novel, LLM-driven experimental framework. This approach simulates a realistic  environment for assessing LLM performance in intent routing, without relying on proprietary IVR data or sensitive customer interactions.

The methodology is structured as a three-stage pipeline, with distinct LLMs assigned to each stage: (1) generating a synthetic IVR menu structure, (2) producing a linguistically diverse set of user intents, and (3) executing natural language intent routing and performance evaluation. This design supports a controlled, reproducible proof-of-concept, allowing us to systematically test the impact of different IVR context representations on routing effectiveness. Table~\ref{tab:llm_roles} summarizes the specific LLMs used in each stage of the experiment. IVR menus and user intents datasets are available at: \url{https://tinyurl.com/Beyond-Touch-Tones-LLMs}. 

\begin{table}[h!]
\centering
\caption{LLM Assignments for Experimental Pipeline}
\label{tab:llm_roles}
\begin{tabular}{lll}
\toprule
\textbf{Id} & \textbf{Model} & \textbf{Role} \\
\midrule
LLM1 & \texttt{gpt-3.5-turbo} & Synthetic IVR menu generation \\
LLM2 & \texttt{gpt-4o-mini} & User intent dataset creation \\
LLM3 & \texttt{gpt-4.1-mini} & Intent interpretation and routing  \\
\bottomrule
\end{tabular}
\end{table}

\subsection{Synthetic IVR Menu Generation}

To simulate a realistic customer service environment, we required a detailed IVR phone menu structure. Due to the proprietary nature of real-world IVRs, LLM1 was prompted to synthesize a 23-node hierarchical menu for a fictitious telecom provider, ``AgentNet''. The prompt instructed LLM1 to emulate the structure of complex enterprise IVRs, covering service areas such as billing, technical support, account management, and new services. The output included structured node descriptions, each detailing its purpose and the available sub-options, with varying depth and complexity across the menu hierarchy.

From this synthetic IVR, we derived two distinct context representations for use with LLM3:

\begin{itemize}
    \item \textbf{Descriptive IVR Menu:} A plain-text, hierarchical outline of the entire IVR structure, preserving all node descriptions and branching logic. This format mimics how a user might comprehend the IVR tree if presented with a full menu listing (see Appendix A).
    
    \item \textbf{Flattened IVR Paths:} A list of all terminal paths expressed as explicit DTMF sequences (e.g., \texttt{1-2-3}), each linked to a specific service destination. This format offers a concise, navigable view of the IVR endpoints, abstracting away intermediate layers (see Appendix B).
\end{itemize}
\subsection{User Intent Dataset Creation}

To evaluate LLM-driven call routing, a diverse and precisely labeled dataset of user intents was essential. Given the absence of public IVR datasets that map free-form complaints to multi-level navigation paths, we employed LLM2 to synthesize and augment such data in a controlled setting with known ground truths. The dataset generation proceeded in two stages:

\begin{enumerate}
    \item \textbf{Base Intent Generation:} LLM2 was prompted to create 10 distinct user complaints or queries for each of the 23 terminal nodes in the AgentNet IVR. These 230 base intents were designed to clearly and uniquely correspond to a specific endpoint, ensuring unambiguous routing paths.

    \item \textbf{Intent Augmentation:} To increase linguistic variability, each base intent was paraphrased into three alternate versions using LLM2. Prompts explicitly instructed the model to preserve semantic equivalence and ground truth mappings while introducing variation in wording, length, and tone. Controlled noise was also induced, including interjections, filler phrases, and minor grammatical deviations, to better simulate real-world inputs.

\end{enumerate}

The final dataset totaled 920 user complaints (230 base + 690 augmented), offering both linguistic diversity and routing precision for evaluating LLM3.

\subsection{Natural Language Intent Routing}

This subsection outlines the experimental setup and prompting strategies used to evaluate LLM3 in natural language call routing. The goal was to assess how effectively LLM3 could map diverse user intents from our synthetic dataset to their correct IVR destinations, given different context formats.

Two experimental conditions were tested, differing only in how the AgentNet IVR menu was presented to LLM3:

\begin{enumerate}
    \item \textbf{Routing with Descriptive IVR Menu:} LLM3 received each user intent alongside the full plain-text hierarchical depiction of the IVR system. Prompts instructed the model to analyze the user query and return the corresponding DTMF path. The template used was:

    \begin{lstlisting}
Given the following IVR menu structure, identify the exact DTMF path (e.g., '1-2-3') that best addresses the user's query. Output only the path.

IVR Menu:
[Full Descriptive IVR Menu Text]

User Query:
[User's natural language intent]
    \end{lstlisting}

    \item \textbf{Routing with Flattened IVR Paths:} In this condition, LLM3 was shown a concise list of terminal paths, each annotated with a brief service description. The prompt instructed the model to select the most appropriate DTMF sequence for the given intent. The template was:

    \begin{lstlisting}
Select the most appropriate DTMF path from the list below that corresponds to the user's query. Output only the path.

Available Paths:
[List of Flattened IVR Paths, e.g., '1-1-1: Check Balance', '1-2-1: Pay Bill']

User Query:
[User's natural language intent]
    \end{lstlisting}
\end{enumerate}

In both conditions, LLM3 processed all 920 user intents. Prompts included strict formatting instructions to ensure the model returned only the DTMF path (e.g., \texttt{1-2-3}) with no additional text. This constraint was essential for automated evaluation. All calls were executed with LLM3 using its default inference settings.

\subsection{Mitigation of Data Contamination Bias}

To prevent data contamination—where evaluation models benefit from prior exposure to task artifacts—we ensured a strict separation of LLM roles throughout the pipeline. The IVR menu was synthesized with help from LLM1 but finalized manually, ensuring a realistic yet unseen structure. User intents and their paraphrases were generated by LLM2, based only on the terminal node labels, with no role in menu creation. Routing was handled by a third model, LLM3, which only received the IVR context in natural language form and had no exposure to prior stages. This separation across generation, augmentation, and evaluation stages minimizes leakage risk and ensures LLM3’s performance reflects genuine generalization.

\subsection{Evaluation Metrics}

LLM3’s routing predictions were evaluated using two metrics. \textbf{Accuracy} measured the proportion of exact matches between predicted and ground truth DTMF paths (e.g., \texttt{1-2-3}). \textbf{Confusion matrices} were used to visualize errors across all 23 terminal paths, highlighting systematic misclassifications. Evaluations were conducted in Python using \texttt{pandas} and \texttt{scikit-learn}.

\section{Results}

This section presents the empirical results of the natural language intent routing experiments conducted with LLM3. We compare routing performance across two key factors: the IVR context format (Descriptive IVR Menu vs. Flattened IVR Paths) and the user intent dataset composition (Base Only vs. Augmented).

\subsection{Overall Routing Accuracy}

The primary evaluation metric was exact-match classification accuracy. Table~\ref{tab:accuracy_results} summarizes LLM3's performance across all four experimental conditions.

\begin{table}[h!]
\centering
\caption{LLM3 Routing Accuracy by IVR Context and Dataset Type}
\label{tab:accuracy_results}
\begin{tabular}{llcc}
\toprule
\textbf{IVR Context} & \textbf{Dataset} & \textbf{Accuracy (\%)} & \textbf{N} \\
\midrule
Flattened Paths & Base Only & 89.13 & 230 \\
Flattened Paths & Augmented & 86.52 & 920 \\
Descriptive Menu & Base Only & 81.30 & 230 \\
Descriptive Menu & Augmented & 77.07 & 920 \\
\bottomrule
\end{tabular}
\end{table}

Across both datasets, LLM3 achieved higher accuracy when provided with the Flattened IVR Paths, with a peak performance of 89.13\% on the Base Only dataset. This indicates that a concise, list-based representation of menu options better supports accurate routing than a verbose hierarchical structure.

In both IVR context types, accuracy was higher on the Base Only dataset compared to its Augmented counterpart. This suggests that while linguistic augmentation introduces useful variation, it may also increase ambiguity or surface edge cases that challenge precise intent classification.

\subsection{Confusion Matrix Insights}

To gain a finer-grained view of LLM3’s routing behavior, confusion matrices were generated for each of the four experimental conditions, capturing classification outcomes across all 23 terminal IVR nodes.

\begin{figure}[h!]
    \centering
    \includegraphics[width=\columnwidth]{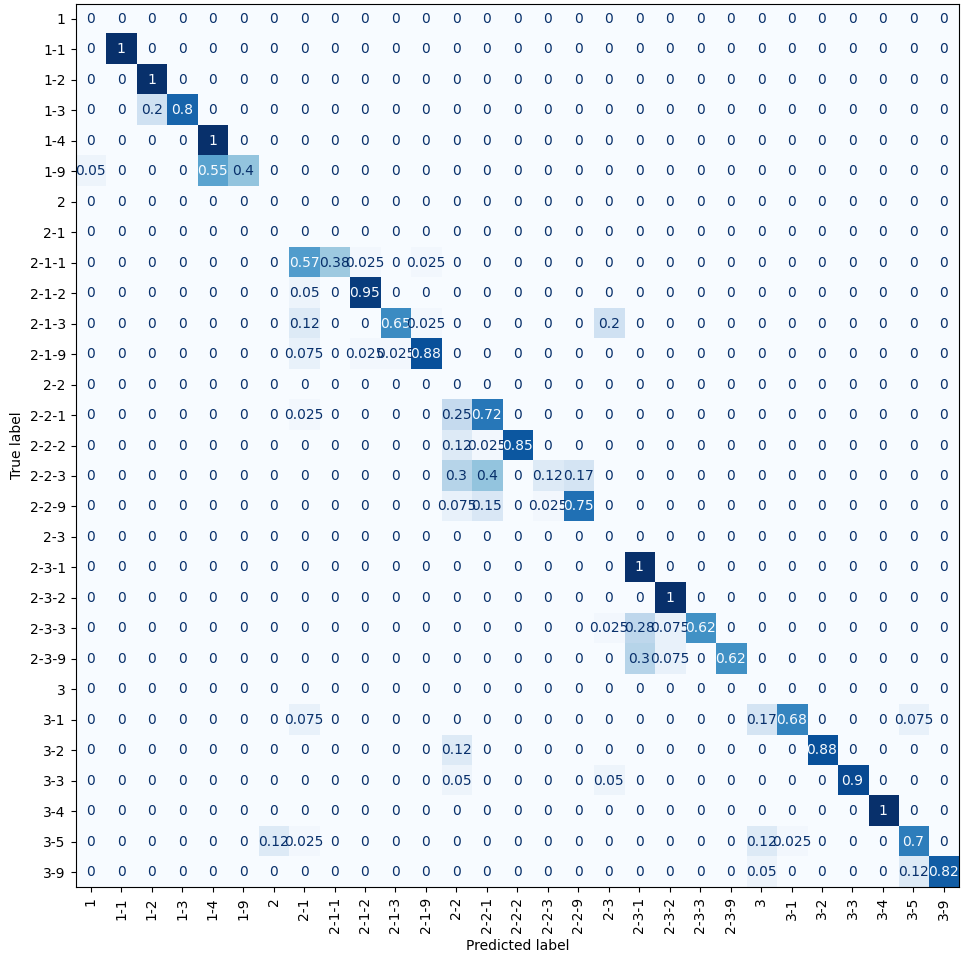}
    \caption{Confusion matrix for Descriptive IVR Menu with Augmented Dataset.}
    \label{fig:cm_full_ivr_augmented}
\end{figure}

\begin{figure}[h!]
    \centering
    \includegraphics[width=\columnwidth]{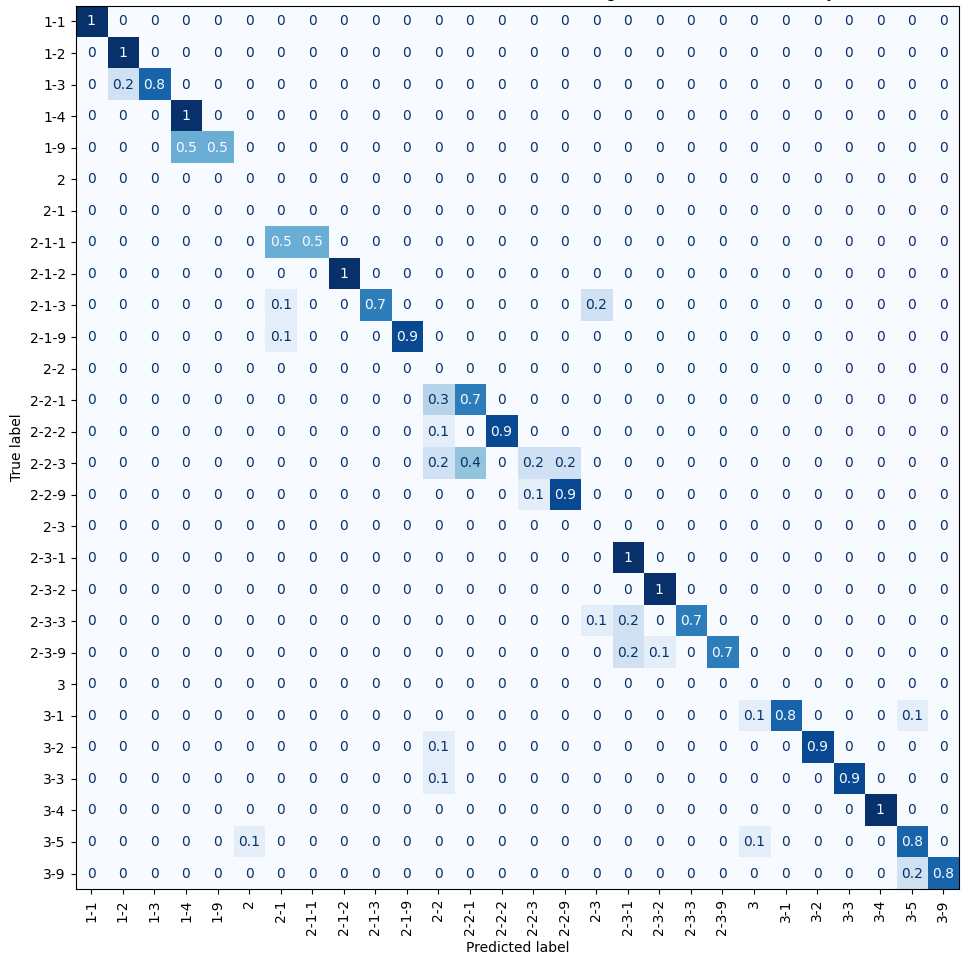}
    \caption{Confusion matrix for Descriptive IVR Menu with Base Only Dataset.}
    \label{fig:cm_full_ivr_base}
\end{figure}

\begin{figure}[h!]
    \centering
    \includegraphics[width=\columnwidth]{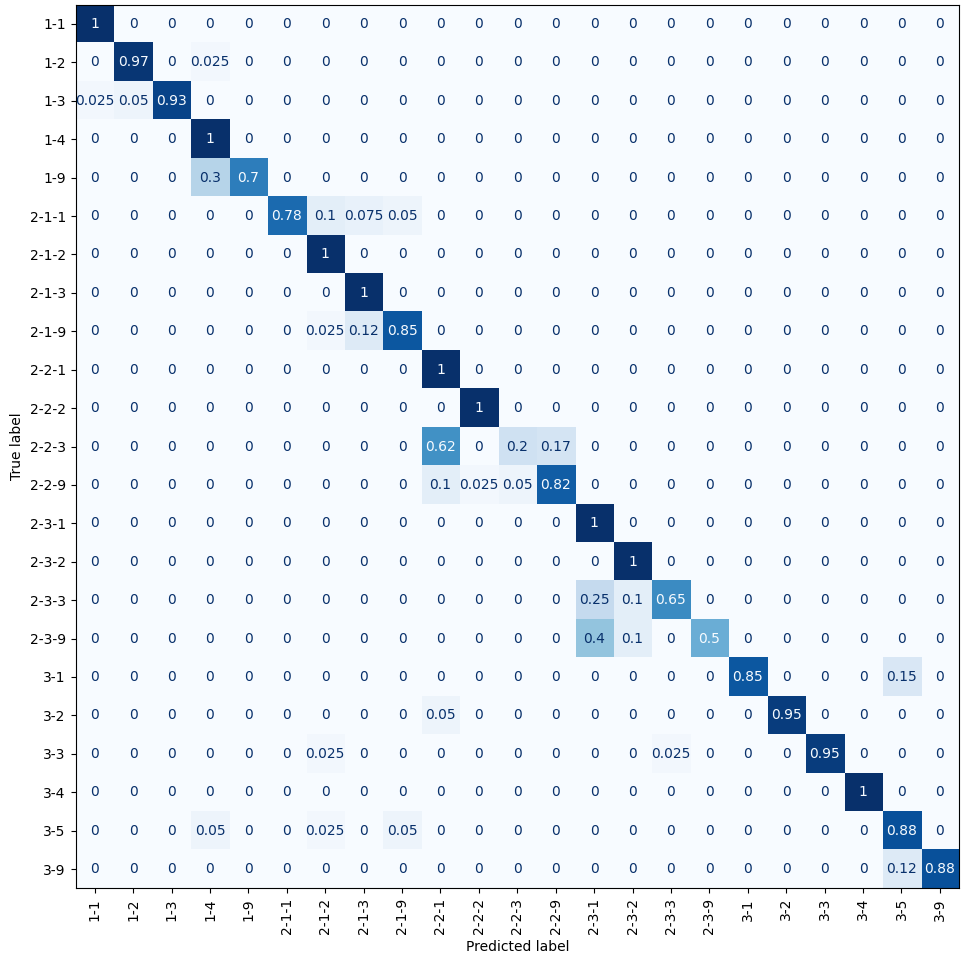}
    \caption{Confusion matrix for Flattened IVR Paths with Augmented Dataset.}
    \label{fig:cm_linear_augmented}
\end{figure}

\begin{figure}[h!]
    \centering
    \includegraphics[width=\columnwidth]{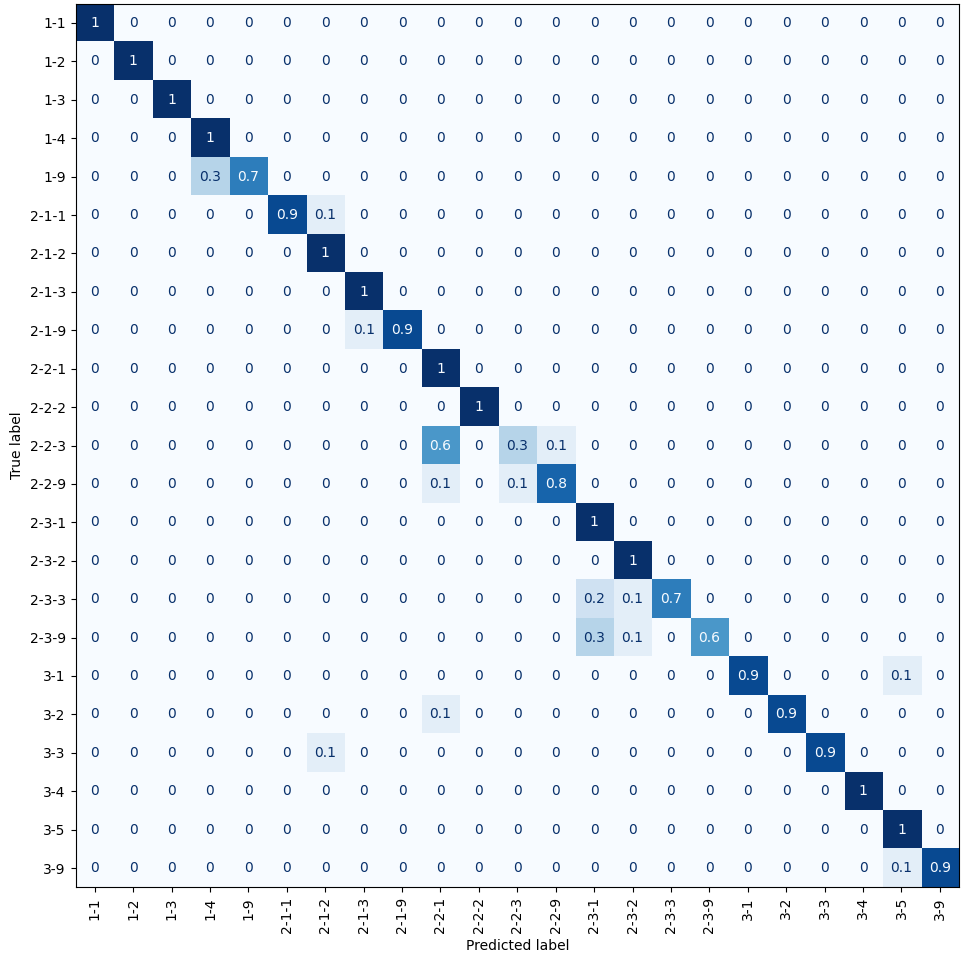}
    \caption{Confusion matrix for Flattened IVR Paths with Base Only Dataset.}
    \label{fig:cm_linear_base}
\end{figure}

Analysis of the matrices (Figures~\ref{fig:cm_full_ivr_augmented}--\ref{fig:cm_linear_base}) reveals consistent patterns in both high-performing and error-prone routing cases.

Several paths exhibited near-perfect F1-scores (e.g., \texttt{1-1}, \texttt{2-1-2}, \texttt{3-4}, \texttt{3-2}, \texttt{3-3}), indicating that LLM3 accurately classified intents corresponding to these frequently encountered or semantically distinctive nodes.

Conversely, certain paths posed consistent challenges. Most notably, \texttt{2-2-3} showed particularly low recall across all conditions—dropping to 0.12 with the Descriptive IVR Menu and 0.20 with the Flattened IVR Paths under the Augmented dataset. This suggests that user intents targeting \texttt{2-2-3} were frequently misclassified, potentially due to semantic ambiguity or overlap with nearby options. Path \texttt{1-9} also exhibited low recall in several settings.

Interestingly, the Flattened IVR Paths improved LLM3’s ability to identify some difficult nodes. For instance, recall for \texttt{2-2-1} increased to 1.00 in the Flattened + Augmented condition, despite lower precision in the Descriptive IVR Menu setting. This indicates that the flattened representation helped the model locate the correct path more reliably, even if it occasionally overpredicted that target.

Overall, these confusion matrix patterns reveal not only the LLM’s strengths in routing clear-cut intents, but also its limitations in disambiguating semantically similar nodes—especially when exposed to linguistically noisy data. The effect of IVR context formatting plays a non-trivial role in mitigating or exacerbating these issues.

\section{Discussion}

\subsection{Key Insights}

Our results show that LLMs route intents more accurately when given flattened IVR paths (up to 89.13\%) than verbose menu descriptions (as low as 77.07\%), suggesting that concise, structured prompts reduce noise and align better with the routing task. This supports the principle that clarity and brevity enhance LLM performance in classification settings. Importantly, transforming menus into flattened paths is a simple, automatable process for real-world use.

Contrary to expectations, performance dropped when using augmented data—despite preserved semantics—likely due to increased lexical variability that made intent mapping noisier. This highlights a trade-off: while augmentation may boost generality, it can reduce precision in tasks requiring exact mappings. Confusion matrix analysis further revealed that certain menu paths consistently underperformed, pointing to ambiguous or overlapping intent definitions. These findings suggest that LLMs can also serve as diagnostic tools for IVR design flaws, such as redundancy or overly granular distinctions that confuse both users and models.

Overall, the study underscores the value of structured prompts and clean data for LLM-based routing, and the potential of LLMs not only as classifiers but as evaluators of IVR menu quality.

\subsection{Limitations and Future Work}

This work used synthetic data and a single IVR structure, limiting generalizability. Real-world phrasings, LLM model and provider diversity, likewise more varied IVRs menues should be explored. We also used exact-match metrics and single-turn routing, whereas real systems often involve near-miss handling and multi-turn interactions. Finally, practical deployment concerns like latency and cost were not assessed and remain important areas for future research.

\section{Conclusion}

This study evaluated how context representation and dataset design affect LLM-based intent routing in IVR systems. We found that concise, flattened representations of IVR paths led to significantly higher accuracy than verbose menu descriptions, and that dataset augmentation—despite increasing linguistic variety—introduced noise that slightly degraded performance. These findings highlight the value of prompt clarity and data precision in LLM applications. Future work should test these patterns on real-world data, explore multi-turn scenarios, and examine how LLMs internally handle structured prompts.

\bibliographystyle{plain}
\bibliography{references}

\begin{thebibliography}{10}

\bibitem{ts2018comparative}
TS. Aswin, Himanshu Batra, and Mathangi Ramachandran.
\newblock Comparative analysis of machine learning algorithms on {IVR} data.
\newblock {\em EAI Endorsed Transactions on Energy Web}, 5(18), 2018.

\bibitem{brown2020language}
Tom~B Brown, Benjamin Mann, Nick Ryder, Melanie Subbiah, Jared Kaplan, Prafulla Dhariwal, Arvind Neelakantan, Pranav Shyam, Girish Sastry, Amanda Askell, et~al.
\newblock Language models are few-shot learners.
\newblock {\em Advances in Neural Information Processing Systems}, 33, 2020.

\bibitem{coman2025ivr}
Ecaterina Coman.
\newblock {IVR} systems used in call center management: a scientometric analysis of the literature.
\newblock {\em Frontiers in Computer Science}, 7, Apr 2025.

\bibitem{ilk2020improving}
N.~Ilk, Guangzhi Shang, and Paulo~B. Góes.
\newblock Improving customer routing in contact centers: An automated triage design based on text analytics.
\newblock {\em Journal of Operations Management}, 66:553--577, Jul 2020.

\bibitem{kosherbay2024ai}
Gassyrbek Kosherbay and Nurgissa Apbaz.
\newblock {AI-Based IVR}.
\newblock {\em arXiv preprint arXiv:2408.10549}, 2024.

\bibitem{openai2023gpt4}
OpenAI.
\newblock {GPT4 Technical Report}.
\newblock {\em arXiv preprint arXiv:2303.08774}, 2023.

\bibitem{ouyang2022training}
Long Ouyang, Jeffrey Wu, Xu~Jiang, Diogo Almeida, Carroll Wainwright, Pamela Mishkin, Chong Zhang, Sandhini Agarwal, Kalian Slama, Alex Ray, and et~al.
\newblock Training language models to follow instructions with human feedback.
\newblock {\em Advances in Neural Information Processing Systems}, 35:27735--27750, 2022.

\bibitem{singh2025ai}
Puneet Singh.
\newblock {AI-Powered IVR and Chat: A New Era in Telecom Troubleshooting}.
\newblock {\em SSRN Electronic Journal}, Jan 2025.

\bibitem{singh2025streamlining}
Puneet Singh.
\newblock {Streamlining Telecom Customer Support with AI-Enhanced IVR and Chat}.
\newblock {\em SSRN Electronic Journal}, Apr 2025.

\bibitem{suhm2002comparative}
Bernhard Suhm, Josh Bers, Dan McCarthy, Barbara Freeman, David Getty, Katherine Godfrey, and Pat Peterson.
\newblock A comparative study of speech in the call center: natural language call routing vs. touch-tone menus.
\newblock In {\em Proceedings of the SIGCHI conference on Human Factors in Computing Systems}, pages 283--290, 2002.

\end{thebibliography}

\appendices

\section*{Appendix A. AgentNet IVR Menu Structure}
\tiny
\fbox{%
\hspace{.5cm}
  \begin{minipage}{0.85\columnwidth}
\ttfamily
IVR Menu Name: "AgentNet IVR"\\

--- Root Menu ---

IVR Message:

  Welcome to FlexiRoute! Connecting you with what you need.
  
  For Billing and Account Management, press 1.
  
  For Technical Assistance, press 2.
  
  For New Services and Upgrades, press 3.
  
  To hear these options again, press 0.
\\ 

--- Branch 1: Billing and Account Management (DTMF: 1) ---

  IVR Message:
  
    You've selected Billing and Account Management.
    
    To check your balance, press 1.
    
    To request your current invoice, press 2.
    
    To make a payment, press 3.
    
    To dispute recent charges, press 4.
    
    To speak with a billing representative, press 9.
    
    To return to the main menu, press 0.  
\\

--- Branch 2: Technical Assistance (DTMF: 2) ---

  IVR Message:
  
    You've selected Technical Assistance. We're here to help.
    
    For Internet service issues, press 1.
    
    For Mobile phone service issues, press 2.
    
    For Video Streaming issues, press 3.
    
    To return to the main menu, press 0.
\\

  --- Sub-Menu for Internet Issues (DTMF: 2-1) ---
  
    IVR Message:
    
      You've selected Internet issues.
      
      To troubleshoot your modem or router, press 1.
      
      For slow internet speeds, press 2.
      
      For service outages, press 3.
      
      To speak with an Internet Technical Representative, press 9.
      
      To return to the previous menu, press 0.      
\\

  --- Sub-Menu for Mobile Phone Service Issues (DTMF: 2-2) ---
  
    IVR Message:
    
      You've selected Mobile phone service issues.
      
      For call and text issues, press 1.
      
      To buy a new phone or upgrade, press 2.
      
      For mobile device support, press 3.
      
      To speak with a Mobile Technical Representative, press 9.
      
      To return to the previous menu, press 0.    
\\

  --- Sub-Menu for Video Streaming Issues (DTMF: 2-3) ---
  
    IVR Message:
    
      You've selected Video Streaming issues.
      
      For quality problems, press 1.
      
      For app or login errors, press 2.
      
      For content or subscriptions, press 3.
      
      To speak with a Video Streaming Technical Representative, press 9.
      
      To return to the previous menu, press 0.
\\

--- Branch 3: New Services and Upgrades (DTMF: 3) ---

  IVR Message:
  
    Welcome to New Services and Upgrades.
    
    To inquire about new Internet plans, press 1.
    
    To inquire about new Mobile phone plans, press 2.
    
    To inquire about Video Streaming packages, press 3.
    
    To add new lines to an existing account, press 4.
    
    To upgrade your current service, press 5.
    
    To speak with a sales representative, press 9.
    
    To return to the main menu, press 0.

  \end{minipage}%
}

\normalsize
\section*{Appendix B. AgentNet IVR Terminal Paths}
\medskip
\tiny
\hspace{-.5cm}
\begin{tabular}{|l|l|l|}
\hline
DTMF & Description & Type \\
Path &  &  \\
\hline
1-1 & Billing and Account Management $\to$ Check Balance & Self-service \\
1-2 & Billing and Account Management $\to$ Request Current Invoice & Self-service \\
1-3 & Billing and Account Management $\to$ Make Payment & Self-service \\
1-4 & Billing and Account Management $\to$ Dispute Recent Charges & Self-service \\
1-9 & Billing and Account Management $\to$ Billing Representative & Agent Handoff \\
2-1-1 & Technical Assistance $\to$ Internet Issues $\to$ Modem / Router Troubleshooting & Self-service \\
2-1-2 & Technical Assistance $\to$ Internet Issues $\to$ Slow Speed & Self-service \\
2-1-3 & Technical Assistance $\to$ Internet Issues $\to$ Service Outages & Self-service \\
2-1-9 & Technical Assistance $\to$ Internet Issues $\to$ Technical Representative & Agent Handoff \\
2-2-1 & Technical Assistance $\to$ Mobile Phone Service Issues $\to$ Connection Issues & Self-service \\
2-2-2 & Technical Assistance $\to$ Mobile Phone Service Issues $\to$ Buy a New Phone & Self-service \\
2-2-3 & Technical Assistance $\to$ Mobile Phone Service Issues $\to$ Mobile Device Support & Self-service \\
2-2-9 & Technical Assistance $\to$ Mobile Phone Service Issues $\to$ Technical Representative & Agent Handoff \\
2-3-1 & Technical Assistance $\to$ Video Streaming Issues $\to$ Quality Problems & Self-service \\
2-3-2 & Technical Assistance $\to$ Video Streaming Issues $\to$ App / Login Errors & Self-service \\
2-3-3 & Technical Assistance $\to$ Video Streaming Issues $\to$ Content / Subscriptions & Self-service \\
2-3-9 & Technical Assistance $\to$ Video Streaming Issues $\to$ Technical Representative & Agent Handoff \\
3-1 & New Services and Upgrades $\to$ New Internet Plans & Self-service \\
3-2 & New Services and Upgrades $\to$ New Mobile Phone Plans & Self-service \\
3-3 & New Services and Upgrades $\to$ New Video Streaming Packages & Self-service \\
3-4 & New Services and Upgrades $\to$ Add New Lines to Existing Account & Agent Handoff \\
3-5 & New Services and Upgrades $\to$ Upgrade Current Service & Agent Handoff \\
3-9 & New Services and Upgrades $\to$ Sales Representative & Agent Handoff \\
\hline
\end{tabular}
\vfill 

\end{document}